\documentclass[11pt,a4paper,final]{article}
\usepackage{amssymb}
\usepackage[final]{graphicx}

\usepackage{newpxtext}
\usepackage[varbb]{newpxmath}
\usepackage{sourcesanspro}

\usepackage{sectsty}
\allsectionsfont{\sffamily}

\usepackage[margin=1in]{geometry}

\usepackage[activate={true,nocompatibility}]{microtype}

\usepackage{enumitem}

\usepackage{setspace}

\usepackage{hyperref}
\hypersetup{bookmarksopen=true,bookmarksnumbered=true,hidelinks,final=true}

\usepackage{doi}

\usepackage[style=numeric, autocite=superscript,
            backend=biber,
						natbib=true,
            sorting=none,
            urldate=comp,
            maxcitenames=2,
            mincitenames=1,
            maxbibnames=3,
            giveninits=true]{biblatex}
\usepackage{array,booktabs}

\usepackage[capitalise,noabbrev]{cleveref}

\usepackage{bm}

\usepackage{mleftright}

\begin{document}

\singlespacing
\thispagestyle{empty}
\begin{titlepage}
	\begin{center}\LARGE\sffamily\bfseries
		Effect modification and non-collapsibility leads to conflicting treatment decisions: a review of marginal and conditional estimands and recommendations for decision-making
	\end{center}
	\vspace{1\baselineskip}
	\begin{raggedright}\footnotesize
		David M.\ Phillippo\footnote{University of Bristol, Canynge Hall, 39 Whatley Road, Bristol, BS8 2PS, UK. Email:	david.phillippo@bristol.ac.uk}\\
		Population Health Sciences, Bristol Medical School, University of Bristol, UK\\[0.5em]
    Antonio Remiro-Azócar\\
		Methods and Outreach, Novo Nordisk Pharma, Madrid, Spain\\[0.5em]
    Anna Heath\\
		Child Health Evaluative Sciences, The Hospital for Sick Children, Toronto, Canada\\
    Dalla Lana School of Public Health, University of Toronto, Toronto, Canada\\
    Department of Statistical Science, University College London, London, United Kingdom\\[0.5em]
    Gianluca Baio\\
		Department of Statistical Science, University College London, London, United Kingdom\\[0.5em]
    Sofia Dias\\
		Centre for Reviews and Dissemination, University of York, York, United Kingdom\\[0.5em]
    A. E. Ades\\
		Population Health Sciences, Bristol Medical School, University of Bristol, UK\\[0.5em]
		Nicky J.\ Welton\\
		Population Health Sciences, Bristol Medical School, University of Bristol, UK
	\end{raggedright}
	\vspace{1\baselineskip}
  \subsection*{Abstract}
  Effect modification occurs when a covariate alters the
  relative effectiveness of treatment compared to control. It is widely
  understood that, when effect modification is present, treatment
  recommendations may vary by population and by subgroups within the
  population. Population-adjustment methods are increasingly used to
  adjust for differences in effect modifiers between study populations and
  to produce population-adjusted estimates in a relevant target population
  for decision-making. It is also widely understood that marginal and
  conditional estimands for non-collapsible effect measures, such as odds
  ratios or hazard ratios, do not in general coincide even without effect
  modification. However, the consequences of both non-collapsibility and
  effect modification together are little-discussed in the literature.

  In this paper, we set out the definitions of conditional and marginal
  estimands, illustrate their properties when effect modification is
  present, and discuss the implications for decision-making. In
  particular, we show that effect modification can result in conflicting
  treatment rankings between conditional and marginal estimates. This is
  because conditional and marginal estimands correspond to different
  decision questions that are no longer aligned when effect modification
  is present. For time-to-event outcomes, the presence of covariates
  implies that marginal hazard ratios are time-varying, and effect
  modification can cause marginal hazard curves to cross. We conclude with
  practical recommendations for decision-making in the presence of effect
  modification, based on pragmatic comparisons of both conditional and
  marginal estimates in the decision target population. Currently,
  multilevel network meta-regression is the only population-adjustment
  method capable of producing both conditional and marginal estimates, in
  any decision target population.

  \paragraph{Keywords} network meta-analysis; population adjustment; effect
  modification; decision-making
\end{titlepage}

\clearpage
\onehalfspacing

\section{Introduction}\label{introduction}

\label{sec:intro}

Healthcare decision makers are frequently tasked with selecting the most
effective treatment from a set of two or more possible candidate
treatments, either purely in terms of treatment effectiveness or as a
balance of cost-effectiveness. This requires reliable estimates of
treatment effects, which are typically obtained from one or more
randomised controlled trials (RCTs). When multiple trials are available,
indirect comparison or network meta-analysis methods are widely used to
synthesise all the evidence in one coherent analysis, even when no
single trial compares all relevant treatments of
interest.\autocite{Bucher1997, Higgins1996, Lu2004, TSD2} Effect modifiers
are factors that alter the relative effectiveness of a treatment
compared to control; for example, if a treatment is more effective for
patients with more severe disease or with certain biomarkers. The
presence of effect modification has strong implications for healthcare
decision-making. First, meta-analyses, indirect comparisons, and network
meta-analyses may be biased if differences in effect modifiers are not
accounted for. Population-adjustment methods such as multilevel network
meta-regression (ML-NMR),\autocite{Phillippo2020_methods} matching-adjusted
indirect comparison (MAIC),\autocite{Signorovitch2010} and simulated
treatment comparison (STC)\autocite{Caro2010, Ishak2015, RemiroAzocar2022}
aim to adjust for differences between study populations using available
individual patient data (IPD) from one or more studies. These methods
are primarily concerned with adjusting for patient characteristics that
may be effect modifiers; study-level effect modifiers related to the
design or context of the trials such as treatment administration or
co-treatments are typically perfectly confounded at the study level and
may require alternative adjustment methods. Second, treatment decisions
may differ between populations or between subgroups within a population,
and so estimates of treatment effects must be produced for the relevant
decision target population (or subgroup thereof). Whilst ML-NMR can
coherently synthesise networks of any size and produce estimates in any
target population, MAIC and STC are limited to pairwise indirect
comparisons between two studies and can only produce estimates relevant
to the population of the aggregate study in the indirect comparison.

For population-level decision making, we are typically interested in
population-average measures of treatment effects, although as we
demonstrate here these may not be sufficient when there is effect
modification. Care is needed to ensure that methods are combining
compatible estimates and to appropriately interpret the results,
particularly when the effect measure of interest is non-collapsible,
such as odds ratios or hazard ratios. A summary effect measure is
non-collapsible when the population-average marginal effects cannot be
expressed as a weighted average of the individual- or subgroup-specific
conditional
effects.\autocite{Gail1984, Greenland1999, Neuhaus1993, Daniel2020} The
result is that conditioning on a covariate that is prognostic of outcome
in the analysis model moves the treatment effect estimate and
fundamentally changes its interpretation, even without interaction or
effect modification.

An estimand defines the exact treatment effect of interest; the
statistical method used to estimate the estimand is an estimator, and
the numerical value computed by the estimator is an estimate. To date,
there has been discussion and disagreement in the literature over
whether population-average conditional or marginal estimands are more
suitable for population-level decision-making based on
effectiveness---including between the authors of this paper. Some of the
authors have previously argued that targeting the conditional estimand
is more desirable, due to increased power to detect treatment effects
and differences, resulting in a more distinct ranking of
treatments;\autocite{Phillippo2020_response_to_RHB} others have previously
argued that the population-average marginal estimand should always be
targeted for population-level decision-making, and that the target
estimand should be selected based on its relevance to the research
question of interest and the decision-making
problem.\autocite{RemiroAzocar2021} However, both of these arguments so far
have not recognised a fundamental issue: the conditional and marginal
estimands correspond to two distinct decision questions that are not
aligned when effect modification is present and may give a different
ranking of treatments.

In this paper, we aim to clarify the different estimands, their
interpretation, and implications for decision making on the basis of
effectiveness and cost-effectiveness. We focus primarily on the context
of population-adjusted indirect comparisons and evidence syntheses,
although the arguments apply equally to any context where
non-collapsible effect measures are used and effect modification is
present, including the analyses of single RCTs and
generalising/transporting RCTs to target populations. We begin by
setting out terminology and defining conditional and marginal estimands
for non-collapsible effect measures with a binary outcome. We describe a
range of current population adjustment approaches and the estimands that
they target. Using a worked example, we then demonstrate the conflict
between population-average conditional and marginal estimands when there
is effect modification, which we interpret in the context of
decision-making. We then make recommendations for decision-making in the
presence of effect modification, before concluding with a discussion.

\section{Defining conditional and marginal
estimands}\label{defining-conditional-and-marginal-estimands}

\label{sec:definitions}

Consider a binary outcome that occurs with event probability
\(\pi_{ik(P)}\) for an individual \(i\) receiving treatment
\(k = A, B, \dots\) in population \(P\) with covariates
\(\bm{x}_{ik(P)}\), under the following model: \begin{equation}
g(\pi_{ik(P)}) = \eta_{k(P)}(\bm{x}_{ik(P)}) = \mu_{(P)} + \bm{m}(\bm{x}_{ik(P)})^{\scriptscriptstyle\mathsf{T}}(\bm{\beta}_1 + \bm{\beta}_{2,k}) + \gamma_k \label{eqn:linpred}
\end{equation} where \(g(\cdot)\) is a link function that transforms
probabilities onto the linear predictor \(\eta_{k(P)}(\cdot)\),
\(\mu_{(P)}\) is the intercept (baseline risk), \(\bm{m}(\cdot)\) is a
function of the covariates, \(\bm{\beta}_1\) are prognostic effects,
\(\bm{\beta}_{2,k}\) are effect-modifying interactions for treatment
\(k\), and \(\gamma_k\) is the individual-level conditional treatment
effect for an individual with \(\bm{x} = \bm{0}\). We set
\(\beta_{2,A} = 0\) and \(\gamma_A = 0\).

Non-collapsibility depends on the choice of link function \(g(\cdot)\),
and in particular on whether the function \begin{equation}
h_{ab}(\pi, \bm{x}) = g^{-1}\mleft(g(\pi) + \bm{m}(\bm{x})^{\scriptscriptstyle\mathsf{T}}(\bm{\beta}_{2,b} - \bm{\beta}_{2,a}) + \gamma_b - \gamma_a\mright)
\end{equation} is linear in \(\pi\).\autocite{Daniel2020, Neuhaus1993} The
function \(h_{ab}(\cdot,\cdot)\), maps the event probabilities on one
treatment \(a\) to event probabilities on another treatment \(b\).
Daniel et al.~\autocite{Daniel2020} term \(h_{ab}(\pi, \bm{x})\) the
characteristic collapsibility function (CCF), and consider the case
where there is no effect modification (\(\bm{\beta}_{2,k} = \bm{0}\) for
all \(k\)) so the CCF no longer depends on the covariates \(\bm{x}\).
When the CCF is not linear in \(\pi\), the corresponding effect measure
is not collapsible; this is the case for example when \(g(\cdot)\) is
the logit or probit link function, which correspond to log odds ratios
or probit differences. When the CCF is linear in \(\pi\), for example
when \(g(\cdot)\) is the identity or log link function, the
corresponding effect measure (risk differences or log risk ratios,
respectively) will be collapsible. However, non-collapsibility is a
necessary consequence of probabilities being bounded between 0 and 1:
modelling collapsible effect measures directly can result in predictions
outside of this range, and induces purely mathematical
treatment-covariate interactions to avoid impossible predictions.

There are several different potential estimands that may be of interest,
and these do not typically coincide for non-collapsible effect measures
such as log odds ratios, even in the absence of effect modification.

The individual-level conditional treatment effects between each pair of
treatments \(b\) vs.~\(a\) for an individual with covariates \(\bm{x}\)
are given by the difference in the linear predictors on each treatment:
\begin{equation}
\begin{aligned}
\label{eqn:individual}
  \gamma_{ab}(\bm{x}) &= \eta_b(\bm{x}) - \eta_a(\bm{x}) \\
    &= \gamma_b - \gamma_a + \bm{m}(\bm{x})^{\scriptscriptstyle\mathsf{T}}(\bm{\beta}_{2,b} - \bm{\beta}_{2,a}).
\end{aligned}
\end{equation} This estimand has an individual-specific interpretation,
provided that relevant sources of subject-level heterogeneity are
accounted for, and depends on the specific values of any effect
modifiers for an individual. However, whilst these individual-level
conditional treatment effects may be of interest to individual patients,
these are not typically the focus for population-level decision making,
which is instead concerned with population-average estimands. A related
estimand is the individual-level conditional effect at the mean
covariate values, \(\gamma_{ab}(\bar{\bm{x}}_{(P)})\), where
\(\bar{\bm{x}}_{(P)}\) is the mean of \(\bm{x}\) in the population
\(P\). Whilst estimates of this estimand are sometimes reported, their
interpretation is problematic, especially with discrete covariates since
it is impossible for an individual to have the ``average'' value; the
individual-level conditional effect at the mean is therefore not useful
for decision-making.

Population-average conditional treatment effects between each pair of
treatments \(a\) and \(b\) in population \(P\) are obtained by averaging
the individual-level treatment effects over the covariate distribution
in the population on the linear predictor scale: \begin{equation}
\begin{aligned}
\label{eqn:conditional}
d_{ab(P)} &= \int_\mathfrak{X} \gamma_{ab}(\bm{x}) \, f_{(P)}(\bm{x}) \mathop{}\!d\bm{x} \\
  &= \int_\mathfrak{X} \bigl(\gamma_b - \gamma_a + \bm{m}(\bm{x})^{\scriptscriptstyle\mathsf{T}}(\bm{\beta}_{2,b} - \bm{\beta}_{2,a}) \bigr) \, f_{(P)}(\bm{x}) \mathop{}\!d\bm{x} 
\end{aligned}
\end{equation} where \(\mathfrak{X}\) is the support of \(\bm{x}\) and
\(f_{(P)}(\bm{x})\) is the joint distribution of \(\bm{x}\) in the
population. The population-average conditional treatment effects
\(d_{ab(P)}\) can be interpreted as the average of the individual-level
treatment effects in the population. Calculating \eqref{eqn:conditional}
requires information on the distribution of effect-modifying covariates
in the population \(P\). In the common special case where
\(\bm{m}(\bm{x}) = \bm{x}\), the covariate means \(\bar{\bm{x}}_{(P)}\)
are sufficient to calculate \eqref{eqn:conditional} since the linear
predictor is linear in the covariates and the integral simplifies to
\(d_{ab(P)} = \gamma_b - \gamma_a + \bar{\bm{x}}^{\scriptscriptstyle\mathsf{T}}_{(P)} (\bm{\beta}_{2,b} - \bm{\beta}_{2,a})\),
where \(\bar{\bm{x}}_{(P)}\) is the mean of \(\bm{x}\) in the population
\(P\).

The individual-level conditional event probabilities on each treatment,
for an individual with covariates \(\bm{x}\), are given by
back-transforming the linear predictor onto the probability scale:
\begin{equation}
\label{eqn:p_individual}
  \pi_{k(P)}(\bm{x}) = g^{-1} \mleft( \eta_{k(P)}(\bm{x}) \mright).
\end{equation} Again, as with the individual-level conditional treatment
effects \(\gamma_{ab(P)}\), the individual-level conditional event
probabilities are relevant to specific individuals and are not typically
the focus for population-level decision making.

Population-average marginal treatment effects are obtained as a summary
of the average event probabilities on each treatment: \begin{equation}
  \Delta_{ab(P)} = g(\bar{\pi}_{b(P)}) - g(\bar{\pi}_{a(P)}) \label{eqn:marginal}
\end{equation} where \begin{equation}
\begin{aligned}
  \bar{\pi}_{k(P)} &= \int_\mathfrak{X} \pi_{k(P)}(\bm{x}) \, f_{(P)}(\bm{x}) \mathop{}\!d\bm{x} \\
    &= \int_\mathfrak{X} g^{-1}(\eta_{k(P)}(\bm{x})) \, f_{(P)}(\bm{x}) \mathop{}\!d\bm{x} \label{eqn:pbar}
\end{aligned}
\end{equation} is the average event probability on each treatment. The
population-average marginal treatment effects \(\Delta_{ab(P)}\) can be
interpreted in terms of the effect of treatment on the average event
probabilities in the population. Calculating \eqref{eqn:p_individual},
\eqref{eqn:marginal}, and \eqref{eqn:pbar} requires information on the
baseline risk \(\mu_{(P)}\) in the population \(P\) (see
\cref{sec:recommendations} for practical considerations);
\eqref{eqn:marginal} and \eqref{eqn:pbar} also require information on
the joint covariate distribution \(f_{(P)}(\bm{x})\) in the population
\(P\).

Here we have defined the estimands in terms of a generative outcome
model. These estimands can also be defined in terms of potential
outcomes without reference to any model; we give definitions in the
potential outcomes framework in \cref{sec:defs_potential_outcomes}. We
note that, although the estimands can be defined in a ``model-free''
manner, all current population adjustment methods with limited IPD will
impose an assumed outcome model in order to form an indirect comparison,
either explicitly (e.g.~STC, ML-NMR) or implicitly
(e.g.~MAIC).\autocite{TSD18}

The terms ``population-average'' and ``marginal'' are often used
interchangeably, but here we make a conceptual distinction.
Population-average refers to a quantity that has been averaged over the
population, which may be conditional (like \(d_{ab(P)}\)) or marginal
(like \(\Delta_{ab(P)}\)), whereas marginal refers to the scale on which
this averaging has taken place (i.e.~the probability scale, rather than
the linear predictor scale for conditional quantities). Similarly, it is
sometimes said that the population-average marginal effect is ``the
effect'' of moving a population from one treatment to another. However,
we clearly see that this intervention effect can be defined in multiple
ways, depending on the scale on which the average is taken. The
population-average marginal estimand \(\Delta_{ab(P)}\) averages the
counterfactual event probabilities on each treatment and compares them,
whereas the population-average conditional estimand \(d_{ab(P)}\)
averages the individual counterfactual treatment effects.

For population-level decision-making, often the decision is made based
on cost-effectiveness rather than purely effectiveness, such as in
health technology assessment. In this case, the estimands above are not
of direct interest, but are instead considered inputs to a
cost-effectiveness model that evaluates the expected net benefit on each
treatment. For now, we consider only effectiveness decisions based on
\(d_{ab(P)}\) or \(\Delta_{ab(P)}\), and we revisit cost-effectiveness
decisions in \cref{sec:interpretation}.

\section{Population adjustment
methods}\label{population-adjustment-methods}

\label{sec:methods}

\begin{table}
  \footnotesize
  \centering
  \caption{Evidence synthesis and population adjustment methods, and the estimands and target populations targeted by each.}
  \label{tab:methods}
  \begin{tabular}{>{\raggedright\let\newline\\\arraybackslash}p{0.2\linewidth}>{\raggedright\let\newline\\\arraybackslash}p{0.3\linewidth}>{\raggedright\let\newline\\\arraybackslash}p{0.3\linewidth}}
\toprule
Method & Estimand & Target population \\
\midrule
(Network) meta-analysis or indirect comparison & Marginal & Average/common population of all included studies \\[1.5em]
IPD network meta-regression & Conditional or marginal & Any \\[1.5em]
ML-NMR & Conditional or marginal & Any \\[0.5em]
MAIC & Marginal & Single aggregate study \\[0.5em]
STC (plug-in means) & Neither, typically combines incompatible conditional and marginal estimates & Single aggregate study \\[1.5em]
STC (simulation) & Marginal & Single aggregate study \\[0.5em]
STC (G-computation) & Marginal & Single aggregate study \\[0.5em]
NMI & Neither, typically combines incompatible conditional and marginal estimates & Any \\
\bottomrule
  \end{tabular}
\\[0.5em]
\footnotesize\raggedright IPD, individual patient data; ML-NMR, multilevel network meta-regression; MAIC, matching-adjusted indirect comparison; STC, simulated treatment comparision; NMI, network meta-interpolation.
\end{table}

In an ideal scenario, IPD would be available from every study, in which
case the ``gold standard'' approach is an IPD network meta-regression
which accounts for differences in effect modifiers between studies and
can produce population-average conditional or marginal treatment effect
estimates in any population of interest (including external target
populations) via equations \eqref{eqn:conditional} and
\eqref{eqn:marginal}.\autocite{Berlin2002, Lambert2002, Riley2010, TSD3}
However, this scenario is uncommon in many practical applications, for
example in health technology assessment where a company making a
submission to an agency such as the National Institute for Health and
Care Excellence in England has IPD from their own study or studies, but
only published aggregate data from their competitors'. Population
adjustment methods are designed with this limited-IPD scenario in mind,
and aim to use IPD available from a subset of studies to account for
differences in the distribution of effect modifiers between
studies.\autocite{TSD18, Phillippo2018} Different population adjustment
methods target different estimands, and produce estimates that are
relevant to different target populations; these are summarised in
\cref{tab:methods}.

ML-NMR is a generalisation of the IPD network meta-regression framework
to incorporate aggregate data, by integrating the individual-level model
over each aggregate study population (as in equation
\eqref{eqn:pbar}).\autocite{Phillippo2020_methods} This approach avoids
aggregation bias, unlike approaches to combining IPD and aggregate data
in network meta-regression that simply ``plug in'' mean covariate values
for the aggregate studies into the individual-level
model.\autocite{Sutton2008, Saramago2012, Donegan2013} ML-NMR combines
evidence at the level of the individual conditional treatment effects,
and can be used to produce estimates of both conditional and marginal
estimands following equations \eqref{eqn:conditional} and
\eqref{eqn:marginal}, in any target population of
interest.\autocite{Phillippo2020_methods, Phillippo2020_response_to_RHB}
The marginalisation integrals \eqref{eqn:pbar} for each aggregate study
are typically calculated using efficient quasi-Monte Carlo numerical
integration, which can also be used to produce estimates for external
target populations with a given covariate
distribution.\autocite{Phillippo2020_methods} The \texttt{multinma} R
package implements ML-NMR models for a range of outcome types, as well
as full-IPD and aggregate data only network meta-analysis as special
cases, and provides functionality to produce estimates of all of the
different estimands defined in \cref{sec:definitions}.\autocite{multinma}

MAIC is a weighting approach, where the method of moments is used to
estimate weights that match covariate means or higher order moments in
an IPD study to those reported in an aggregate
study.\autocite{Signorovitch2010} MAIC targets the population-average
marginal estimand \eqref{eqn:marginal}, but since no conditional
regression model is fitted this bypasses the need to evaluate any
marginalisation integral (equation \eqref{eqn:pbar}) entirely. MAIC can
only produce estimates relevant to the aggregate study population in a
two-study indirect comparison.\autocite{TSD18}

STC is a regression adjustment approach that fits a regression model in
an IPD study and uses this to predict outcomes in an aggregate study
population.\autocite{Caro2010} The most common form of STC typically
combines a conditional estimate from the IPD study, obtained by
plugging-in mean covariate values to equation \eqref{eqn:individual},
with a marginal estimate as reported by the aggregate study, which are
incompatible and is thus biased against both the population-average
conditional \eqref{eqn:conditional} and marginal \eqref{eqn:marginal}
estimands.\autocite{RemiroAzocar2021, Phillippo2020_response_to_RHB} Other
forms of STC are available that avoid this problem and target the
population-average marginal estimand via simulation (to evaluate
equations \eqref{eqn:marginal} and \eqref{eqn:pbar}), however these
incur additional sampling variation by trying to simulate a limited
number of participants in the aggregate trial.\autocite{Caro2010} All forms
of STC can only produce estimates relevant to the aggregate study
population in a two-study indirect comparison.\autocite{TSD18}

A more sophisticated form of STC based on G-computation addresses
several of the issues with other forms of STC.\autocite{RemiroAzocar2022}
G-computation STC targets the population-average marginal estimand
\eqref{eqn:marginal}, fitting a regression model in an IPD study, and
evaluating the marginalisation integral \eqref{eqn:pbar} over an
aggregate study population using simulation (parametric G-computation).
Uncertainty is fully quantified by implementing the approach in a
Bayesian framework.\autocite{RemiroAzocar2022} However, like other forms of
STC, this approach can only produce estimates relevant to the aggregate
study population in a two-study indirect
comparison.\autocite{RemiroAzocar2022, TSD18}

Population adjustment analyses are often required to make simplifying
assumptions for identifiability, most commonly invoking the \emph{shared
effect modifier assumption}, which states that effect modifier
interactions are equal for a set of treatments; that is
\(\bm{\beta}_{2,k} = \bm{\beta}_{2,\mathcal{T}}\) for all treatments
\(k\) in the set \(\mathcal{T}\). This assumption may be reasonable if
treatments in \(\mathcal{T}\) are from the same class and share a mode
of action, otherwise this assumption is unlikely to hold.\autocite{TSD18}
For MAIC and STC, with a continuous outcome and a linear outcome model,
making the shared effect modifier assumption for treatments \(B\) and
\(C\) means that the estimated relative treatment effect
\(\Delta_{BC(AC)}\) is constant across populations, and can be applied
in any target population and not just the \(AC\) trial population. For
other outcomes and outcome models, however, this assumption is not
sufficient to make the marginal effect \(\Delta_{BC(AC)}\)
transportable, as this marginal effect is specific to the distribution
of covariates and baseline risk in the \(AC\) population. For ML-NMR,
the shared effect modifier assumption may be used in smaller networks to
identify the model in the absence of sufficient data, for example in a
two-study indirect comparison.\autocite{Phillippo2020_methods} ML-NMR can
use this assumption regardless of the outcome type or outcome model,
since it is applied to the individual-level conditional outcome model.
If this assumption does not hold and the model cannot be identified via
other means (e.g.~external information to inform prior distributions for
the interactions or other structural assumptions about interactions),
then ML-NMR is limited to producing estimates in the aggregate study
population, like MAIC and STC. In even moderately-sized networks, ML-NMR
may allow the shared effect modifier assumption to be assessed or
removed entirely.\autocite{Phillippo2022_multinomial} We examine the shared
effect modifier assumption further in \cref{sec:ex_sharedEM}.

Network meta-interpolation (NMI) is different to other regression-based
approaches; whilst an outcome regression model is defined, this model is
not estimated directly. Instead, published subgroup analyses from each
trial are ``interpolated'' to a specific target population by solving
equations based on the best linear unbiased predictor, and then these
adjusted estimates are combined in a standard NMA.\autocite{Harari2023} The
motivation of NMI is to produce population-adjusted estimates without
making the shared effect modifier assumption, by using additional
information in the form of subgroup analyses from all
studies.\autocite{TSD18, Phillippo2020_methods} However, NMI incurs similar
biases to plug-in means STC, as the reported study-level treatment
effect estimates in trial publications are unlikely to be compatible
with the conditional treatment effect estimates required by NMI to
perform interpolation. NMI therefore typically mixes incompatible
estimates within the model (within studies, as opposed to across studies
for plug-in means STC), and is thus biased against both the conditional
and marginal estimands. Furthermore, since the outcome regression model
is not fully estimated, it is not possible for NMI to produce
population-average marginal treatment effects or absolute predictions
(e.g.~average event probabilities). In certain specific scenarios
(i.e.~binary covariates and linear outcome models) NMI does appear
promising; however, the properties and performance of NMI are not yet
fully understood, and simulation studies have not directly investigated
bias or adequacy of variance estimation.

Standard network meta-analysis, pairwise meta-analysis, and indirect
comparison methods combine aggregate data from each study without
adjusting for covariates, targeting a population-average marginal
estimand. The target population is an average of the included study
populations, which are typically assumed to all be representative of a
single common population.

\section{Example}\label{example}

\label{sec:example}

To illustrate the issues that arise with non-collapsible effect measures
when there is effect modification, we consider a simple example with a
single covariate \(x\) that is both prognostic of outcome and effect
modifying, uniformly distributed in the population between \(-1\) and
\(1\), and binary outcomes on three treatments \(k=A,B,C\) that are
modelled on the log odds ratio scale using \cref{eqn:linpred} with the
logit link function \(g(\pi) = \mathop{\mathrm{logit}}(\pi)\) and
\(m(x) = x\).

\subsection{\texorpdfstring{Effect modification can result in
conflicting rankings
\label{sec:ex_EM_ranks}}{Effect modification can result in conflicting rankings }}\label{effect-modification-can-result-in-conflicting-rankings}

It is well-understood that the magnitudes of population-average
conditional and marginal effects \(d_{ab(P)}\) and \(\Delta_{ab(P)}\)
will typically differ, even in the absence of effect
modification.\autocite{Gail1984, Greenland1999, Neuhaus1993, Daniel2020}
However, when there is effect modification the direction of effect can
also differ between \(d_{ab(P)}\) and \(\Delta_{ab(P)}\), resulting in
conflicting treatment rankings. It is straightforward to find values of
the coefficients where this occurs, for example \(\mu_{(P)} = 0\),
\(\beta_1 = -1\), \(\beta_{2,B} = -3\), \(\beta_{2,C} = -1\),
\(\gamma_B = -4\), \(\gamma_C = -3\). This results in population-average
conditional treatment effects \(d_{AB(P)} = -4\) and \(d_{AC(P)} = -3\)
(i.e.~\(B\) better than \(C\) for reducing a harmful outcome), but
population-average marginal treatment effects \(\Delta_{AB(P)} = -2.36\)
and \(\Delta_{AC(P)} = -2.49\) (i.e.~\(C\) better than \(B\)). This is
because treatment \(C\) results in a lower average event probability
overall: \(\bar{\pi}_{B(P)} = 0.087\), \(\bar{\pi}_{C(P)} = 0.077\) (and
\(\bar{\pi}_{A(P)} = 0.5\)).

\begin{figure}
\centering
\includegraphics{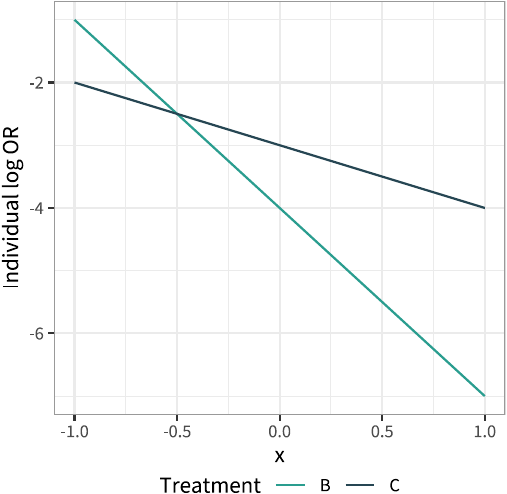}
\caption{Individual-level log odds ratios \(\gamma_{AB}(x)\) and
\(\gamma_{AC}(x)\) for treatments \(B\) and \(C\) compared to \(A\),
over the range of the covariate \(x\) in the
population.\label{fig:individual_logOR}}
\end{figure}

Basing a decision on these population-average marginal treatment effects
would result in choosing treatment \(C\), but 75\% of the population are
given an inferior treatment and are expected to do better on treatment
\(B\) (\cref{fig:individual_logOR}). On the other hand, basing a
decision on these population-average conditional treatment effects would
result in choosing treatment \(B\), but a lower expected number of
events \(N\bar{\pi}_{k(P)}\) (in a population of size \(N\)) would be
achieved by treatment \(C\).

\subsection{\texorpdfstring{Marginal ranks depend on baseline risk and
prognostic factors
\label{sec:ex_marginal_and_bl_risk}}{Marginal ranks depend on baseline risk and prognostic factors }}\label{marginal-ranks-depend-on-baseline-risk-and-prognostic-factors}

\begin{figure}
\centering
\includegraphics{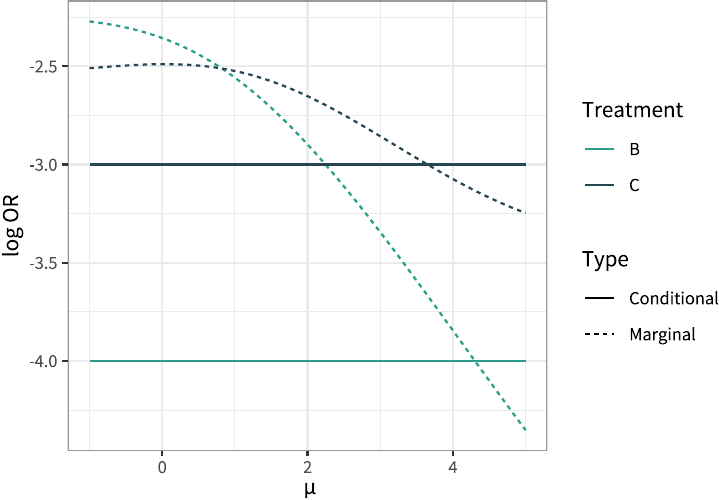}
\caption{Population-average marginal (\(\Delta_{AB(P)}\) and
\(\Delta_{AC(P)}\)) and conditional (\(d_{AB(P)}\) and \(d_{AC(P)}\))
log odds ratios for treatments \(B\) and \(C\) compared to \(A\), for a
range of values of the baseline risk
\(\mu_{(P)}\).\label{fig:marginal_logOR_blrisk}}
\end{figure}

Furthermore, the population-average marginal treatment effects and
rankings are dependent on the baseline risk \(\mu_{(P)}\), as well as
the distribution of the covariate \(x\) (even if it is only prognostic),
but the population-average conditional treatment effects and rankings
only change when the distribution of effect-modifying covariates
changes. \Cref{fig:marginal_logOR_blrisk} shows the population-average
conditional and marginal treatment effects for a range of values of the
baseline risk \(\mu_{(P)}\). The population-average conditional
treatment effects \(d_{ab(P)}\) are constant over all values of the
baseline risk, but the population-average marginal treatment effects
\(\Delta_{ab(P)}\) change depending on the baseline risk and switch
ranks. Also note that the widely-known result that conditional effects
are always further from the null than marginal effects
\autocite{Neuhaus1993, Daniel2020} does \emph{not} hold when there is
effect modification; here the marginal log odds ratio for treatment
\(C\) is further from the null for \(\mu_{(P)} \ge 4.4\). Consequently,
the corollary result that power is always greater for conditional
effects also does not hold when there is effect modification.

\begin{figure}
\centering
\includegraphics{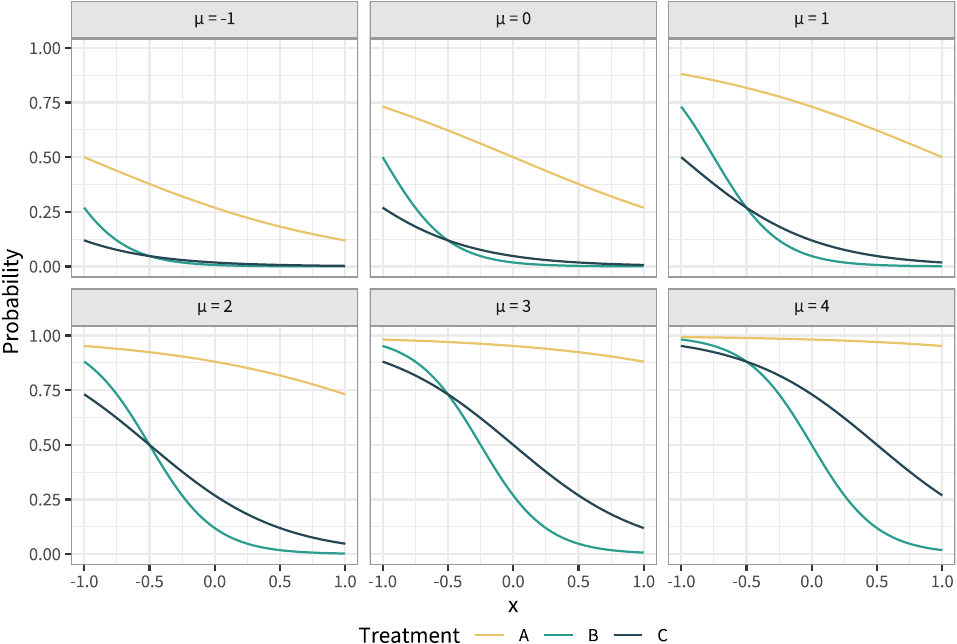}
\caption{Individual event probabilities \(\pi_{k(P)}(x)\) on each
treatment over the range of covariate values \(x\) in the population,
for a range of values of the baseline risk
\(\mu_{(P)}\).\label{fig:event_probs_blrisk}}
\end{figure}

The population-average marginal effects change because changing the
baseline risk changes the individual event probabilities
(\cref{fig:event_probs_blrisk}), which are averaged over the population
to obtain \(\bar{\pi}_{k(P)}\) and the marginal effects.

\subsection{\texorpdfstring{The shared effect modifier assumption
\label{sec:ex_sharedEM}}{The shared effect modifier assumption }}\label{the-shared-effect-modifier-assumption}

\begin{figure}
\centering
\includegraphics{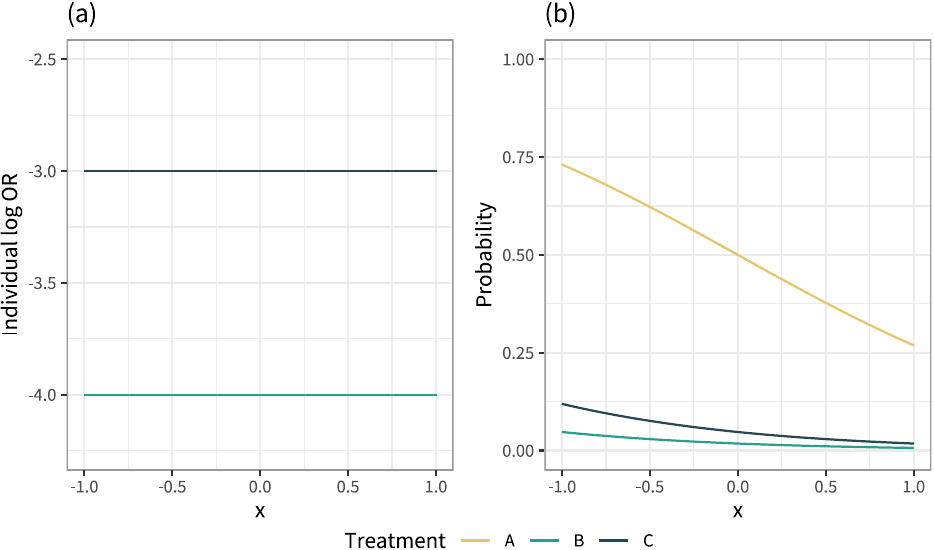}
\caption{Individual-level log odds ratios \(\gamma_{Ak}(x)\) (a) and
event probabilities \(\pi_{k(P)}(x)\) (b) when there is no effect
modification (\(\beta_{2,B} = \beta_{2,C} = 0\)), over the range of the
covariate \(x\) in the population.\label{fig:no_EM}}
\end{figure}

We see in \cref{fig:event_probs_blrisk} that the curves of
individual-level event probabilities \(\pi_{k(P)}(x)\) by covariate
\(x\) on each treatment intersect. This crossing of event probability
curves due to effect modification is the reason that the conditional and
marginal treatment effects can give different rankings. If there is no
effect modification then the individual-level log odds ratios between
all treatments are constant (\cref{fig:no_EM}a), the event probability
curves cannot cross (\cref{fig:no_EM}b), and the conditional and
marginal rankings and decision questions are aligned.

\begin{figure}
\centering
\includegraphics{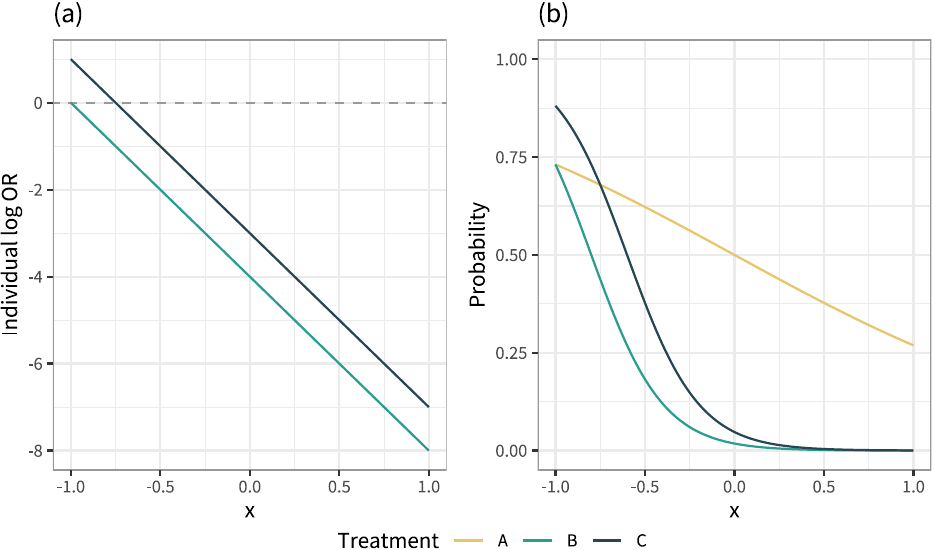}
\caption{Individual-level log odds ratios \(\gamma_{Ak}(x)\) (a) and
event probabilities \(\pi_{k(P)}(x)\) (b) when the shared effect
modifier assumption is made for treatments \(B\) and \(C\)
(\(\beta_{2,B} = \beta_{2,C} = -4\)), over the range of the covariate
\(x\) in the population.\label{fig:shared_EM}}
\end{figure}

Moreover, the individual event probability curves of two treatments
cannot cross if the effect modifier interaction coefficients are the
same for these two treatments; for example, if
\(\beta_{2,B} = \beta_{2,C}\) (\cref{fig:shared_EM}b). In this situation
the individual-level odds ratio between these two treatments is again
constant (the curves in \cref{fig:shared_EM}a are parallel). The
assumption that effect modifier interaction coefficients are the same
for a set of treatments is called the shared effect modifier assumption
(see \cref{sec:methods}).\autocite{TSD18, Phillippo2018} The shared effect
modifier assumption may sometimes be used for ML-NMR when there are
insufficient data to estimate separate interaction terms for each
treatment, for example in a two-study indirect comparison, in order to
produce estimates for populations other than the aggregate study
population.\autocite{Phillippo2020_methods} Pairs of treatments between
which the shared effect modifier assumption is not made can have
individual odds ratios and event probability curves that cross; in
\cref{fig:shared_EM}a the individual odds ratio for \(C\) (compared to
\(A\)) intersects the line of no effect, and as a result the event
probability curves for \(A\) and \(C\) cross. This means that, even if
the shared effect modifier assumption is made for treatments \(B\) and
\(C\), the marginal ranking of treatment \(A\) can still change. With a
continuous outcome and linear outcome model, the shared effect modifier
assumption can also be used by MAIC and STC in order to produce relative
effect estimates that are relevant to a target population other than
that of the aggregate study in the indirect comparison, since then
\(\Delta_{BC(P)} = d_{BC(P)}\) is constant across all
populations.\autocite{TSD18} In all other cases, the population-average
marginal treatment effects \(\Delta_{BC(AC)}\) produced by MAIC and STC
(with simulation) are specific to the distributions of covariates and
baseline risk in the aggregate \(AC\) study population, and cannot be
transported to another population with different distributions of
covariates or baseline risk.

\subsection{\texorpdfstring{An example with a binary covariate in a
contingency table
\label{sec:ex_contingency}}{An example with a binary covariate in a contingency table }}\label{an-example-with-a-binary-covariate-in-a-contingency-table}

To help further solidify these ideas, we now consider an example with a
binary outcome and a single binary covariate through a contingency
table. Consider a trial of four treatments \(A\), \(B\), \(C\), and
\(D\), randomised equally, in a population where the prevalence of a
biomarker \(x\) is 25\%. This biomarker is prognostic and
effect-modifying, and we are interested in reducing occurrence of some
harmful event. \Cref{tab:ex_contingency} shows the numbers of
individuals who did and did not experience the event, within subgroups
defined by the biomarker \(x\) and over the whole population.

\begin{table}
  \footnotesize
  \centering
  \caption{Contingency table for an illustrative example of four treatments, stratified by a biomarker $x$. Population-average marginal odds ratios, subgroup-specific conditional odds ratios, and population-average conditional odds ratios are calculated vs. treatment $A$.}
  \label{tab:ex_contingency}
  \begin{tabular}{lrrrrrr}
\toprule
& \multicolumn{2}{c}{$x = 0$} & \multicolumn{2}{c}{$x = 1$} & \multicolumn{2}{c}{Overall population} \\
\cmidrule(lr){2-3} \cmidrule(lr){4-5} \cmidrule(lr){6-7}
Treatment & Events & Non-events & Events & Non-events & Events & Non-events \\
\midrule
$A$ & 202 & 548 & 156 &  94 & 358 & 642 \\[1em]

$B$ &  89 & 661 &  42 & 208 & 131 & 869 \\[0.5em]
\multicolumn{1}{r}{Marginal OR} & & & & & \multicolumn{2}{r}{$(131/869)/(358/642) = 0.27$} \\
\multicolumn{1}{r}{Conditional OR} & \multicolumn{2}{r}{$(89/661)/(202/548) = 0.37$} & %
  \multicolumn{2}{r}{$(42/208)/(156/94) = 0.12$} & %
  \multicolumn{2}{r}{$\exp(0.75 \times \log(0.37) + 0.25 \times \log(0.12)) = 0.28$} \\[1em]

$C$ &  13 & 737 & 144 & 106 & 157 & 843 \\[0.5em]
\multicolumn{1}{r}{Marginal OR} & & & & & \multicolumn{2}{r}{$(157/843)/(358/642) = 0.33$} \\
\multicolumn{1}{r}{Conditional OR} & \multicolumn{2}{r}{$(13/737)/(202/548) = 0.05$} & %
  \multicolumn{2}{r}{$(144/106)/(156/94) = 0.82$} & %
  \multicolumn{2}{r}{$\exp(0.75 \times \log(0.05) + 0.25 \times \log(0.82)) = 0.10$} \\[1em]

$D$ & 137 & 613 &   6 & 244 & 143 & 857 \\[0.5em]
\multicolumn{1}{r}{Marginal OR} & & & & & \multicolumn{2}{r}{$(143/857)/(358/642) = 0.30$} \\
\multicolumn{1}{r}{Conditional OR} & \multicolumn{2}{r}{$(137/613)/(202/548) = 0.61$} & %
  \multicolumn{2}{r}{$(6/244)/(156/94) = 0.01$} & %
  \multicolumn{2}{r}{$\exp(0.75 \times \log(0.61) + 0.25 \times \log(0.01)) = 0.24$} \\
\bottomrule
  \end{tabular}
\end{table}

In \cref{tab:ex_contingency}, we then calculate the population-average
marginal odds ratios, subgroup-specific conditional odds ratios, and
population-average conditional odds ratios for each treatment compared
to \(A\). Here, since we have a binary covariate, calculating the
population-average conditional odds ratios via the integral in equation
\eqref{eqn:conditional} simplifies to taking the weighted average (on
the log odds ratio scale) of the subgroup-specific conditional odds
ratios according to the prevalence of the biomarker in the population.
Based on the population-average marginal odds ratios, treatment \(B\) is
the best, as it results in the lowest number of events overall. However,
the population-average conditional odds ratios give a different ranking:
treatments \(C\) and \(D\) are both ranked better than \(B\), with
treatment \(C\) being the best.

Treatment \(C\) is the most effective treatment for most individuals in
this population, i.e.~in the biomarker-negative subgroup (\(x=0\)) which
makes up 75\% of the population. This leads to \(C\) having the best
population-average conditional odds ratio. However, \(C\) is less
effective than \(B\) for the smaller biomarker-positive subgroup
(\(x=1\)); the increased number of events in this subgroup result in a
higher number of events overall on treatment \(C\) than \(B\), and hence
a worse population-average marginal odds ratio.

Treatment \(D\) is less effective than \(B\) for most of this
population---the biomarker-negative subgroup (\(x=0\))---and has a
higher event rate overall. The population-average marginal odds ratio
for treatment \(D\) is therefore worse than for treatment \(B\).
However, \(D\) is highly effective for the biomarker-positive subgroup
(\(x=1\)), to an extent that is sufficient to give a better
population-average conditional odds ratio than treatment \(B\).

We see here how the population-average conditional and marginal effects
weigh up effectiveness over the population differently. The
population-average marginal effects weigh up the expected number of
events overall, i.e.~the average is taken on the probability scale. The
population-average conditional effects weigh up the expected individual
or subgroup effectiveness over the population, i.e.~the average is taken
on the additive linear predictor scale.

As shown earlier in \cref{sec:ex_EM_ranks}, we again see here that the
treatment with the best population-average marginal effect (\(B\)) is
not always the best treatment for the majority of individuals when
effect modification is present. Furthermore, when covariates are
discrete or non-symmetrically distributed, or treatment or covariate
effects are non-linear, the treatment with the best population-average
conditional effect is not always the best treatment for the majority of
individuals. In this example with a binary covariate, treatment \(D\)
has a better population-average conditional effect than treatment \(B\),
but \(D\) is less effective than \(B\) for most individuals.

Selecting the single treatment \(B\) with the best population-average
marginal effect results in a decision that minimises the number of
events overall. However, the rank conflict with the population-average
conditional effects indicates that there is substantive differential
effectiveness within the population, and a decision stratified by
subgroup may be closer to optimal. In this case, treatment \(B\) is
inferior for every individual in the population: treating
biomarker-negative individuals with \(C\) and biomarker-positive
individuals with \(D\) is the optimal decision. This stratified
treatment decision would result in the least number of events overall, a
population-average marginal odds ratio of \((19/981)/(358/642) = 0.03\),
and a population-average conditional odds ratio of
\(\exp(0.75 \times \log(0.05) + 0.25 \times \log(0.01)) = 0.04\).

\section{Interpretation}\label{interpretation}

\label{sec:interpretation}

Rank-switching between the population-average conditional and marginal
effects can only occur in the presence of effect modification, between
treatments that have different interaction terms (i.e.~no shared effect
modifier assumption), and when this causes treatment ranks to change
across individuals/subgroups in the population. We give a formal proof
of this statement in \cref{sec:em_proof}. However to see this
intuitively, consider that rankings based on the population-average
marginal treatment effects \(\Delta_{ab(P)}\) can only change compared
to the population-average conditional treatment effects \(d_{ab(P)}\) if
the individual event probabilities on each treatment
\(\pi_{k(P)}(\bm{x})\) change ranks within the population (i.e.~if the
event probability curves cross as in \cref{fig:event_probs_blrisk}).
This can happen if and only if the individual-level treatment effects
\(\gamma_{Ak}(\bm{x})\) change ranks within the population (i.e.~if the
individual treatment effect curves cross as in
\cref{fig:individual_logOR}), which can happen if and only if there is
effect modification between the two treatments.

It is well-understood that population-average marginal treatment effects
are population-specific, and depend on the distributions of baseline
risk and prognostic factors, as well as any effect modifiers. However,
when there is effect modification we have seen that the marginal
treatment rankings can also change---even if the only factors that
change are those that do not affect individual treatment effects
(baseline risk, prognostic factors). Conditional population-average
treatment effects and rankings will only change depending on the
distribution of effect modifiers, and do not depend on baseline risk or
prognostic factors.

The population-average conditional and marginal effects have different
interpretations, and correspond to different decision questions
regarding effectiveness. The population-average conditional treatment
effects represent the average of the individual treatment effects
experienced in the population. They answer the question ``Which
treatment has the greatest effect for individuals, on average, in this
population?'' The population-average marginal treatment effects (whether
odds ratios, risk ratios, or risk differences) quantify effectiveness in
terms of the average event probabilities on each treatment, and for
non-collapsible effect measures, by definition, do not represent any
average of the individual or subgroup treatment effects experienced in
the population, even without effect modification. They answer the
question ``Which treatment minimises (or maximises) the marginal average
event probability in this population?''

These two decision questions are equivalent \emph{except} when there is
effect modification. When there is effect modification and the
individual treatment effects cross within the population, the two can
give different rankings. As a result, a decision based on
population-average marginal treatment effects to obtain the minimum (or
maximum) average event probability can result in choosing a treatment
that is inferior for the majority of individuals in the population.
Conversely, a decision based on population-average conditional treatment
effects to obtain the most effective treatment on average for each
individual in the population can result in choosing a treatment that
gives a higher (or lower) average event probability than another
treatment option.

Either of these decision questions and estimands might be justified.
Basing a decision on the population-average conditional effects means
that decision-makers want to maximise the benefit, on average, for each
individual in the population. Basing a decision on the
population-average marginal effects means that decision-makers want to
minimise (or maximise) the expected number of events over the
population, for example if (non-)events have a high associated cost or
disutility. Indeed, in many cases decision-makers may wish to satisfy
both decision questions; however, when there is effect modification
there may not be a single treatment that achieves this over the entire
population.

For cost-effectiveness decisions, neither of the above decision
questions or estimands are of direct interest. Instead, the decision
question is ``Which treatment maximises the expected net benefit in this
population?'', and the relevant quantity is the expected net benefit
(NB) on each treatment \(\mathbb{E}(\mathrm{NB}_{k(P)})\). The net
benefit is typically a function \(\varphi_{k(P)}(\cdot, \cdot)\) of the
average event probabilities \(\bar{\pi}_{k(P)}\), or the individual
event probabilities \(\pi_{k(P)}(\bm{x})\) for the distribution of the
covariates in the population, as well as other parameters
\(\bm{\theta}_{k(P)}(\bm{x})\) such as resource use costs and adverse
events which may also vary over the population. When there is patient
heterogeneity, such as that caused by effect modification, a
cost-effectiveness analysis should handle this appropriately by
averaging (integrating) net benefit over the population, which
necessitates constructing the net benefit as a function of the
individual event probabilities \(\pi_{k(P)}(\bm{x})\):\autocite{Welton2015}
\begin{equation}\label{eqn:netbenefit}
\mathrm{NB}_{k(P)} = \int_\mathfrak{X} \varphi_{k(P)} \mleft( \pi_{k(P)}(\bm{x}), \bm{\theta}_k(\bm{x}) \mright) \, f(\bm{x}) \mathop{}\!d\bm{x}.
\end{equation} In simple cases this integral may be evaluated directly,
however discrete event simulation is often used instead to construct and
evaluate such cost-effectiveness models.\autocite{Karnon2012, Karnon2014}
Comparing equation \eqref{eqn:netbenefit} with equations
\eqref{eqn:conditional} and \eqref{eqn:pbar}, we see that the
population-average conditional estimand corresponds to a net benefit
function that is linear in individual treatment effects
\(\varphi_{k(P)}(\bm{x}) = \gamma_{Ak}(\bm{x})\) and the
population-average marginal estimand corresponds to a net benefit
function that is linear in individual event probabilities
\(\varphi_{k(P)}(\bm{x}) = \pi_{k(P)}(\bm{x})\), when effectiveness is
the only consideration.

\section{Considerations for survival
outcomes}\label{considerations-for-survival-outcomes}

\label{sec:survival}

Similar arguments can be applied to survival or time-to-event outcomes
analysed using proportional hazards models, since hazard ratios are also
non-collapsible. Consider a general proportional hazards model where the
hazard function for an individual \(i\) receiving treatment \(k\) in
population \(P\) with covariates \(\bm{x}_{ik(P)}\) at time \(t\) is
\begin{subequations}\label{eqn:ph_model}
\begin{align}
  h_{k(P)}(t \mid \bm{x}_{ik(P)}) &= h_{0(P)}(t) \exp\mleft( \eta_{k(P)}(\bm{x}_{ik(P)}) \mright) \\
  \eta_{k(P)}(\bm{x}_{ik(P)}) &= \mu_{(P)} + \bm{x}^{\scriptscriptstyle\mathsf{T}}_{ik(P)} (\bm{\beta}_1 + \bm{\beta}_{2,k}) + \gamma_k
\end{align}
\end{subequations} for some baseline hazard function \(h_{0(P)}(t)\),
with corresponding survival function
\(S_{k(P)}(t\mid\bm{x}_{ik(P)}) = \exp( -\int^t_0 h(u) \mathop{}\!du )\).

The population-average marginal hazard ratio \(\Delta_{ab(P)}(t)\) is
\begin{equation}\label{eqn:marginal_hr}
  \Delta_{ab(P)}(t) = \frac{ \bar{h}_{b(P)}(t) }{ \bar{h}_{a(P)}(t) },
\end{equation} the ratio of the marginal hazard functions
\begin{equation}\label{eqn:marginal_hazard}
  \bar{h}_{k(P)}(t) = \frac{ \int_\mathfrak{X} S_{k(P)}(t \mid \bm{x}) \, h_{k(P)}(t \mid \bm{x}) \, f_{(P)}(\bm{x}) \mathop{}\!d\bm{x} }{ \bar{S}_{k(P)}(t) },
\end{equation} where \(\bar{S}_{k(P)}(t)\) is the population-average
marginal survival function (or standardised survival function)
\begin{equation}\label{eqn:marginal_survival}
  \bar{S}_{k(P)}(t) = \int_\mathfrak{X} S_{k(P)}(t \mid \bm{x}) \, f_{(P)}(\bm{x}) \mathop{}\!d\bm{x}.
\end{equation} The marginal hazard function \(\bar{h}_{k(P)}(t)\) can be
considered a weighted average of the individual-level hazard functions,
weighted by the probability of surviving to time \(t\). Notably, the
population-average marginal hazard ratio \(\Delta_{ab(P)}(t)\) is
\emph{time-varying}, as well as depending on the shape of the baseline
hazard function \(h_{0(P)}(t)\), and the distributions of baseline
hazard and all prognostic and effect modifying covariates. This means
that, if covariates are present, proportional hazards mathematically
cannot hold at the marginal level. This is the case whether the
covariates are prognostic or effect modifying; however, an argument
analogous to that in \cref{sec:em_proof} shows that effect modification
is necessary for the marginal hazard functions to cross (assuming that
covariates are balanced between arms at baseline, and that the same
baseline hazard function \(h_{0(P)}(t)\) applies in both arms).

The population-average conditional log hazard ratio \(d_{ab(P)}\) under
this model is again given by equation \eqref{eqn:conditional}---the
average treatment effect over the population, taken on the log hazard
scale. Again, \(d_{ab(P)}\) depends only on the distribution of effect
modifiers, not on the shape of the baseline hazard function, or the
distributions of baseline hazard or prognostic factors, and is not
time-varying.

\begin{figure}
\centering
\includegraphics{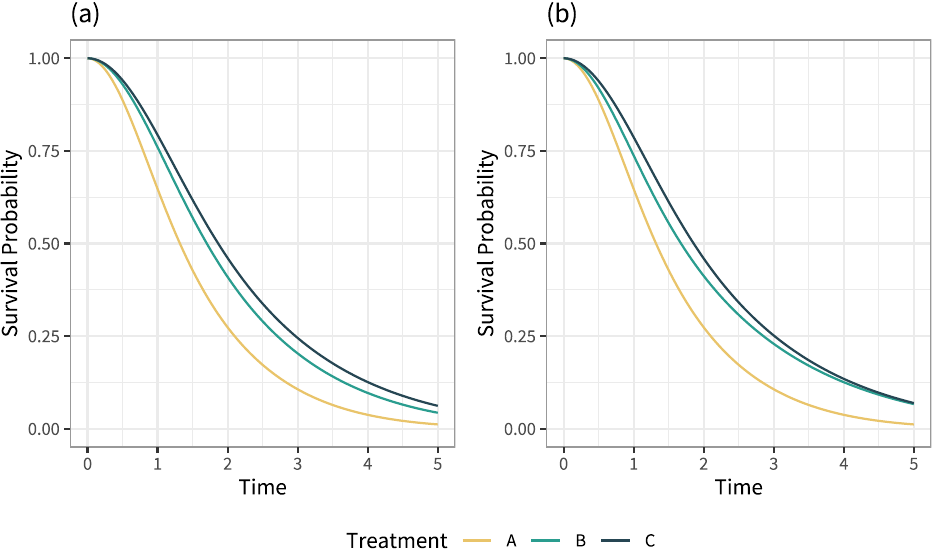}
\caption{Population-average marginal survival curves with a single
uniformly-distributed covariate that is (a) prognostic only, (b)
prognostic and effect modifying.\label{fig:weibull_surv}}
\end{figure}

As an example, consider a Weibull proportional hazards model for three
treatments, \(A\), \(B\), and \(C\), with a single covariate \(x\) that
is uniformly distributed in the population between \(-1\) and \(1\),
with survival and hazard functions \begin{subequations}
\begin{align}
    S_{k(P)}(t \mid x) &= \exp\mleft( -t^{\nu_{(P)}} \exp(\eta_{k(P)}(x)) \mright)\\
    h_{k(P)}(t \mid x) &= \nu_{(P)}  t^{\nu_{(P)} - 1} \exp(\eta_{k(P)}(x))
\end{align}
\end{subequations} For simplicity, we use a common shape parameter
\(\nu_{(P)} = 2\) for all three treatments, and set \(\mu_{(P)} = -1\),
\(\gamma_B = \log(0.6)\), and \(\gamma_C = \log(0.5)\). The covariate
\(x\) we set to be prognostic of survival with \(\beta_1 = \log(0.25)\).
We then consider two scenarios, one where \(x\) is only prognostic so
\(\beta_{2,B} = \beta_{2_C} = 0\), and the other where \(x\) is
moderately effect modifying, \(\beta_{2,B} = \log(0.7)\) and
\(\beta_{2,C} = \log(0.9)\). The population-average marginal survival
curves under this set-up are shown in \cref{fig:weibull_surv}.

\begin{figure}
\centering
\includegraphics{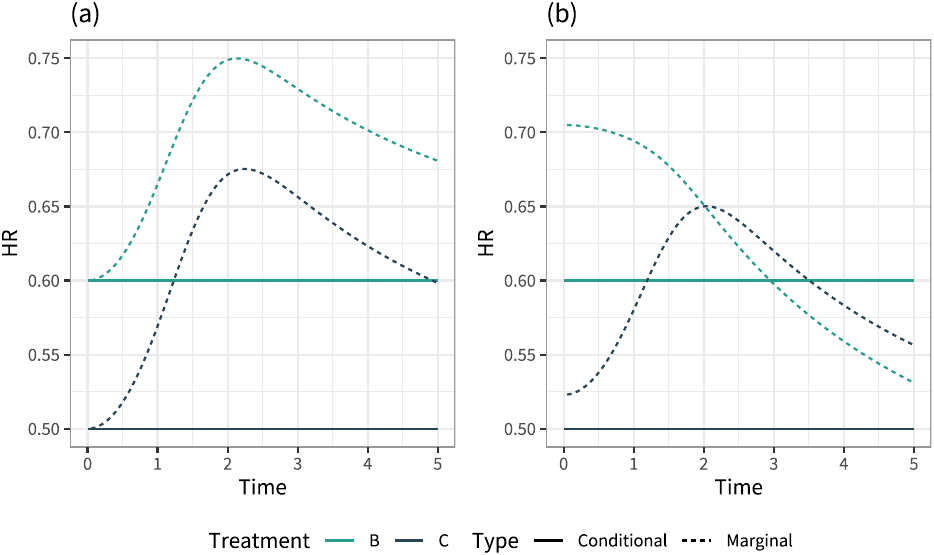}
\caption{Population-average conditional and marginal hazard ratios
vs.~treatment \(A\) over time with a single uniformly-distributed
covariate that is (a) prognostic only, (b) prognostic and effect
modifying.\label{fig:weibull_hrs}}
\end{figure}

\begin{figure}
\centering
\includegraphics{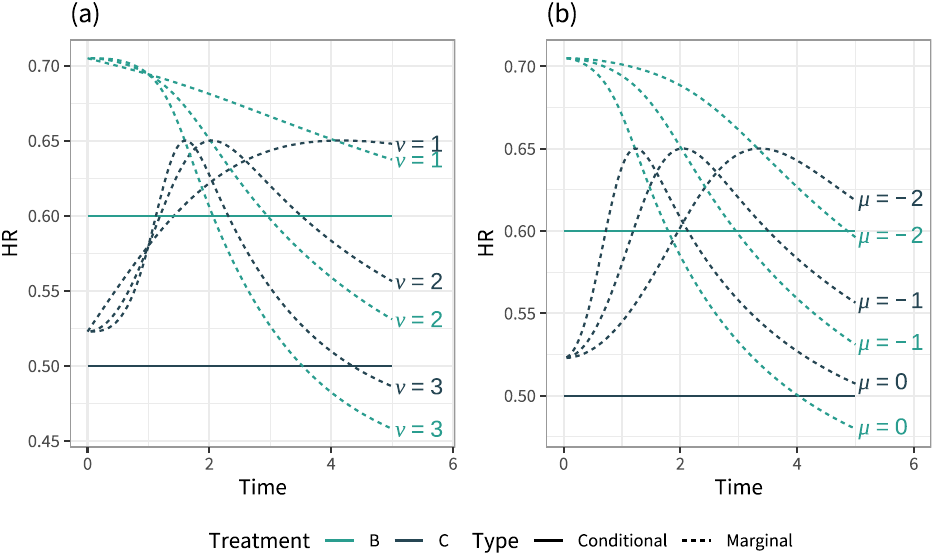}
\caption{Population-average conditional and marginal hazard ratios
vs.~treatment \(A\) over time, varying (a) the shape of the baseline
hazard function \(\nu_{(P)}\), and (b) the distribution of baseline log
hazard \(\mu_{(P)}\).\label{fig:weibull_hrs_bhaz}}
\end{figure}

The corresponding population-average conditional and marginal hazard
ratios are shown in \cref{fig:weibull_hrs}. The presence of a prognostic
covariate means that the population-average marginal hazard ratios are
no longer constant over time; proportional hazards does not hold at the
marginal level. When this covariate is also effect modifying, the
population-average marginal hazard ratios can also change ranks over
time. \Cref{fig:weibull_hrs_bhaz} demonstrates that the
population-average marginal hazard ratios depend on the shape of the
baseline hazard function and the distribution of baseline hazard,
whereas the population-average conditional hazard ratios do not.

With survival outcomes, the key quantities for decision-making are
typically the population-average marginal survival functions
\(\bar{S}_{k(P)}(t)\) on each treatment and summaries thereof, such as
survival probabilities at clinically relevant time points, median
survival times, or (restricted) mean survival times.
\(\bar{S}_{k(P)}(t)\) is also the typically the primary input to an
economic model for decisions based on cost-effectiveness.

\section{Recommendations for decision-making in the presence of effect
modification}\label{recommendations-for-decision-making-in-the-presence-of-effect-modification}

\label{sec:recommendations}

Decision-makers should specify \emph{a priori} the target population and
decision question that are of interest, and analysts should ensure that
the corresponding estimand is appropriately targeted---be that
population-average conditional or marginal estimates for effectiveness
decisions, or the necessary inputs to an economic model for
cost-effectiveness decisions. In a health technology assessment context,
guidance for submissions to the National Institute of Health and Care
Excellence (NICE) in England states that the choice of effect modifiers
must be pre-specified prior to analysis, clinically plausible, and
justified through empirical evidence, expert opinion, or systematic
review.\autocite{NICE_Manual2022, TSD18} Similarly, section 4.9 of the NICE
health technology evaluation manual expresses a preference for
pre-specified identification of subgroups with biological plausibility,
and warns against post-hoc ``dredging'' for subgroup
effects.\autocite{NICE_Manual2022} This applies to subgroups based both on
effectiveness (i.e.~effect modifiers), and on other factors such as
costs, baseline risk, or adverse events.

\subsection{Decisions based on effectiveness
only}\label{decisions-based-on-effectiveness-only}

For population decision-making based purely on effectiveness, the
relevant estimands are population-average conditional or marginal
treatment effects in the decision target population, for example after
population adjustment. ML-NMR can produce relevant estimates for any
decision target population;\autocite{Phillippo2020_methods} other
approaches like MAIC and STC are limited to producing estimates in the
population of the aggregate study in an indirect comparison, which may
not represent the decision target population.\autocite{TSD18} In small
networks like a two-study indirect comparison, ML-NMR may make use of an
identifying assumption such as the shared effect modifier assumption to
produce estimates for populations other than the aggregate study
population (\cref{sec:methods}). MAIC and STC cannot make use of the
shared effect modifier assumption except for continuous outcomes with a
linear outcome model, and are restricted to producing estimates for the
aggregate study population. Only ML-NMR at present can produce estimates
of both conditional and marginal estimands. The population-average
conditional and marginal estimands can result in conflicting rankings
when there is effect modification, and in some cases a decision-maker
may be forced to choose between a treatment that is inferior for the
majority of the population or one that results in a worse expected
number of events overall. If this is a concern, then the only way to
resolve this conflict and realign the two effectiveness decision
questions and estimands is to allow different decisions within subgroups
based on covariate values.

With just one effect modifier, it is straightforward to visualise the
impact on decisions by plotting the individual treatment effects against
the covariate (as in \cref{fig:individual_logOR}), but with multiple
effect modifiers this quickly becomes infeasible. It is possible to
mathematically determine the boundaries between different optimal
treatment choices in the \(L\)-dimensional covariate space, where \(L\)
is the number of effect-modifying covariates. However, this is likely to
result in complex decisions that may be difficult to implement and hard
to justify.

We propose a more pragmatic approach, where decision-makers consider
both the population-average conditional and marginal treatment effects.
If the respective rankings agree, then there is no substantial conflict
to resolve and a single decision might be justified for the entire
population. This does not rule out the possibility that smaller
subgroups might obtain greater treatment benefit or lower (or higher)
average event probabilities from a different decision, but it does mean
that the overall decision is both the most effective on average for each
individual in the population and results in the lowest (or highest)
average event probability overall. If the rankings are in conflict, then
decision-makers could attempt to resolve this by considering making
decisions for subgroups formed by a small number of effect modifiers.
These chosen effect modifiers should be those that have the most impact
on treatment effects in the population, based both on the strength of
the interaction and on the range of covariate values in the population.
The precise cut-points could be based on examining plots of the
individual treatment effects against the effect modifiers in question
one covariate at a time, holding the other covariates at the population
means, or they could be guided by clinical reasoning or practice (say if
there are established thresholds for normal vs.~high values). Careful
selection of subgroups in this manner is likely to be sufficient to
resolve the conflict between decision questions, whilst keeping the
resulting subgroup decisions simple enough for decision-makers to
justify and implement.

When effect modification is not present, the two decision questions are
aligned and both estimands will give the same ranking of treatments. In
a Bayesian setting Bayesian \(p\)-values and the precision of the ranks
will be identical between the population-average conditional and
marginal effects, since the action of marginalisation is a monotonic
transformation of the posterior about the origin.

\subsection{Decisions based on
cost-effectiveness}\label{decisions-based-on-cost-effectiveness}

For decisions based on cost-effectiveness, patient heterogeneity (such
as that caused by effect modification) should be handled appropriately
by averaging net benefit over the population as in equation
\eqref{eqn:netbenefit}.\autocite{Welton2015} Discrete event simulation
based on individual or subgroup event probabilities or survival curves
is one suitable approach,\autocite{Karnon2012, Karnon2014} however such
models can be complex to develop and evaluate. As a result, discrete
event simulation is not as widely used as other approaches such as
Markov models or decision trees, which typically do not account for
patient heterogeneity and are based on average event probabilities
\(\mathrm{NB}_{k(P)} = \varphi_{k(P)}(\bar{\pi}_{k(P)}, \bm{\theta}_{k(P)})\)
or average survival curves
\(\mathrm{NB}_{k(P)} = \varphi_{k(P)}(\bar{S}_{k(P)}, \bm{\theta}_{k(P)})\).
Assuming that the structure of \(\varphi_{k(P)}(\cdot, \cdot)\) is
maintained between approaches and that any additional parameters
\(\bm{\theta}_{k(P)}(\bm{x})\) are averaged over the population in the
same way, how different the resulting expected net benefit is depends on
the non-linearity of \(\varphi_{k(P)}(\cdot, \cdot)\) with respect to
\(\pi_{k(P)}(\bm{x})\) (due to Jensen's inequality), with equality if
\(\varphi_{k(P)}(\cdot, \cdot)\) is linear in \(\pi_{k(P)}(\bm{x})\).
However, there are likely to also be structural differences in
\(\varphi_{k(P)}(\cdot, \cdot)\) between approaches which may introduce
further differences between the results.

Regardless of the cost-effectiveness model chosen, care must be taken to
ensure that the relevant inputs are produced appropriately. In many
cases these are the event probabilities on each treatment in the
decision target population, either average \(\bar{\pi}_{k(P)}\) or
individual \(\pi_{k(P)}(\bm{x})\). These should be obtained by applying
relevant relative treatment effect estimates (e.g.~after
population-adjustment into the target population) to a representative
distribution for baseline risk. Evidence for this baseline risk
distribution need not be obtained from the trial(s) used to estimate
relative effects; indeed this may ideally be obtained from a
representative registry or cohort study in the decision target
population.\autocite{TSD5} Typically this baseline risk distribution is on
the average event probability \(\bar{\pi}_{k_0(P)}\) for a given
treatment \(k_0\).

To then obtain average event probabilities on all other treatments, one
simple approach is to rearrange equation \eqref{eqn:marginal} as
\(\bar{\pi}_{k(P)} = g^{-1}\mleft(g(\bar{\pi}_{k_0(P)}) + \Delta_{k_0 k(P)}\mright)\)
and apply population-average marginal treatment effect estimates to the
baseline risk distribution. However, this is only correct if the
population-average marginal effects \(\Delta_{ab(P)}\) were produced by
marginalising over the same baseline risk distribution, since these are
not transportable across populations with different baseline risks (or
covariate distributions). This means that population adjustment methods
like MAIC and STC that can only estimate population-average marginal
treatment effects in the aggregate study population of an indirect
comparison are strictly limited to producing average event probabilities
in this aggregate study population; the population-average marginal
estimates are specific to this population, and cannot be applied to
another population with a different distribution of baseline risk even
if the covariate distributions are the same.

A more sophisticated approach when a regression model has been used is
to instead apply equation \eqref{eqn:pbar}, averaging absolute model
predictions on each treatment over the target
population.\autocite{Phillippo2020_methods} This requires that model
\eqref{eqn:linpred} is estimated for all treatments, which is currently
only possible using ML-NMR in a population adjustment setting. This also
requires a distribution on the intercept \(\mu_{(P)}\) in the target
population, instead of the baseline risk \(\bar{\pi}_{k_0(P)}\); a
procedure for converting between the two by solving equation
\eqref{eqn:pbar} for values of \(\mu_{(P)}\) given a sample of values
for \(\bar{\pi}_{k_0(P)}\) is given in Phillippo et
al.\autocite{Phillippo2022_multinomial} Using this approach, average event
probabilities can be produced in any target population of interest; an
example using ML-NMR is given in Phillippo et
al.\autocite{Phillippo2022_multinomial} In small networks like a two-study
indirect comparison, ML-NMR may make use of an identifying assumption
such as the shared effect modifier assumption to produce estimates for
populations other than the aggregate study population
(\cref{sec:methods}).

The same considerations apply for cost-effectiveness decisions involving
survival outcomes. Furthermore, since the presence of covariate
effects---the very motivation for performing population
adjustment---implies that marginal hazard ratios are time-varying, the
oft-used practice of applying a constant hazard ratio for each treatment
to a baseline survival function in the economic model is not appropriate
here. Instead, the estimated individual- or subgroup-specific survival
curves \(S_{k(P)}(t \mid \bm{x})\) or population-average survival curves
\(\bar{S}_{k(P)}(t)\) should be used directly in the modelling. MAIC and
STC can obtain population-average survival curves on all treatments, but
only in the population of the aggregate study. ML-NMR can estimate
either individual-level or population-average survival curves on all
treatments, in any population.\autocite{Phillippo2024_survival} As well as
information on the covariate distribution in the target population, a
distribution for the baseline hazard is required. When Kaplan-Meier data
are available in the target population, even on a single treatment arm,
this can be used to estimate the appropriate baseline hazard parameters
within the ML-NMR model\autocite{Phillippo2024_survival}; otherwise,
baseline hazard estimates could be taken from a study in the network
where the baseline hazard is deemed to be representative, whilst still
allowing for differences in covariate distributions to be accounted for.

Carrying out a cost-effectiveness analysis does not resolve the fact
that making a single treatment decision for an entire population may be
sub-optimal, and that greater cost-effectiveness might be obtained by
allowing different decisions for subgroups.

\section{Discussion}\label{discussion}

\label{sec:discussion}

In this paper, we have argued that population-average conditional and
marginal estimands correspond to different decision questions, either to
maximise average effectiveness or to minimise (or maximise) average
event probabilities respectively. We have demonstrated how the presence
of effect modification means that these estimands and decision questions
are no longer aligned, and may not correspond to the same ranking of
treatments. Moreover, making a single treatment decision in the presence
of effect modification can result in either selecting an inferior
treatment for the majority of individuals, or selecting a treatment with
a worse average event probability overall. Where allowable, making
decisions by subgroups may result in patients being given a more
effective treatment for them and result in greater cost-effectiveness
overall. However, identification of valid subgroups is non-trivial;
analyses to detect interactions and subgroups typically have low power,
there is a risk of spurious findings particularly if many candidate
factors are considered, and precision will be reduced within subgroups
which may weaken conclusions.

We have focused the motivation for this paper on population-level
decision making using population-adjusted analyses like MAIC or ML-NMR,
which typically involve two or more trials and three or more
treatments.\autocite{TSD18} However, our arguments apply equally to
analyses of single trials, and where there are only two treatments
(\cref{sec:ex_sharedEM}). Whilst trials do not typically consider
adjusting for effect modifiers as a primary analysis, consideration of
effect modifiers is central to generalising or transporting treatment
effects to target populations (typically a marginal estimand is targeted
through a propensity score analysis);\autocite{Cole2010, Stuart2011} such
analyses are therefore subject to exactly the same issues that we
describe here.

In the two-study indirect comparison scenario for which MAIC and STC are
proposed, MAIC and STC (with simulation or G-computation) can produce
estimates of population-average marginal treatment effects relevant to
the aggregate study population. However, these methods cannot produce
estimates for another target population of interest, except in the
special case of continuous outcomes and a linear model when the shared
effect modifier assumption may be applied (\cref{sec:methods}). In the
two-study indirect comparison scenario, ML-NMR can produce both
conditional and marginal estimates relevant to the aggregate study
population, and can produce estimates relevant to any target population
of interest with the use of an additional identifying assumption such as
the shared effect modifier assumption (\cref{sec:methods}). In larger
networks, which may often be available in practice,\autocite{Phillippo2019}
ML-NMR may allow the shared effect modifier assumption to be assessed or
avoided entirely.\autocite{Phillippo2022_multinomial} IPD meta-regression
is the ideal---but uncommon---special case of ML-NMR where IPD are
available from every study, in which case there are sufficient data to
estimate the model without additional identifying assumptions. To be
relevant for decision-making, whichever method is used, estimates must
be produced that are relevant to the decision target
population.\autocite{TSD18}

Previous discussion of non-collapsibility, such as the excellent paper
by Daniel et al.,\autocite{Daniel2020} has largely focused on scenarios
where there is no effect modification. In such cases, there are
well-known results that i) conditional estimands lie further from the
null than marginal estimands; ii) conditional estimators may have
reduced precision compared to marginal estimators; iii) the reduction in
precision is outweighed by the increased separation from the null,
resulting in increased power for conditional estimands (as summarised by
Daniel et al.\autocite{Daniel2020}). Whilst comparing precision of
population-average marginal and conditional estimators is not a
meaningful comparison of like with like, this power comparison is
meaningful because the two estimands share the same null. However, we
have demonstrated that these results break down when there is effect
modification, as the population-average marginal estimand is no longer
always closer to the null---even if the distribution of effect modifiers
is unchanged (\cref{sec:ex_marginal_and_bl_risk}).

Whilst for binary outcomes we focused on log odds ratios with a logit
link function, the same issues and arguments apply to other
non-collapsible effect measures such as the summary effect from a probit
link model. The same arguments also apply to non-collapsible effect
measures for other types of outcomes. For example, when analysing
ordered categorical outcomes using ordered logistic or probit
regression, there is a single summary population-average conditional
treatment effect across all outcome categories, but the
population-average marginal treatment effects differ for each category
and the marginal rankings may change between
categories.\autocite{Phillippo2022_multinomial} This should not be
construed as a reason to prefer modelling collapsible effect measures
such as risk differences or log risk ratios directly, e.g.~by the use of
an identity or log link function with a binary outome. Such models can
result in predicted probabilities less than 0 or greater than 1, and
will induce purely mathematical treatment-covariate interactions.

For survival or time-to-event outcomes analysed using (log) hazard
ratios, we have seen that not only do population-average marginal hazard
ratios depend on the shape of the hazard function and the distribution
of baseline hazard and all prognostic and effect modifying covariates,
but they must also vary over time. Crucially this means that, whenever
covariates are present, proportional hazards \emph{cannot} hold at the
marginal level. Such covariates do not need to be effect modifying or
time-varying; even with purely prognostic baseline covariates the
mathematical consequence is that the population-average marginal hazard
ratio is time-varying. A positive consequence of this, however, is that
adjustment for covariates measured only at baseline can be sufficient to
address violations of proportional hazards in unadjusted models (a
phenomenon we have noted previously \autocite{Phillippo2024_survival}).
Daniel et al.\autocite{Daniel2020} also considered the implications of
non-collapsibility of the hazard ratio. They considered a marginal
hazard ratio that had been additionally marginalised over time, as well
as the covariates and baseline hazard, to provide a single
non-time-varying marginal hazard ratio, and proposed an approach to
obtain such a hazard ratio from an adjusted model. Daniel et al.~note
that such marginal hazard ratios (and thus any marginal rankings based
on them) are further dependent on the length of study follow-up and
observed censoring pattern, as well as the shape of the baseline hazard,
and distributions of baseline risk and prognostic and effect modifying
covariates.

For cost-effectiveness decisions, when effect modification or other
sources of patient heterogeneity are present, the net benefit should be
averaged over the population.\autocite{Welton2015} This requires the
production of individual- or subgroup-specific event probabilities; in
the context of population-adjusted indirect comparisons or evidence
syntheses of multiple studies, at present this is only possible using
ML-NMR.\autocite{Phillippo2020_methods} Discrete event simulation
\autocite{Karnon2012, Karnon2014} is one suitable approach to constructing
a net benefit function and averaging this over a population, however it
is not widely used due to complexity. Markov models and decision trees
are much more prevalent, however these approaches do not typically
account for patient heterogeneity. Determining the possible extent of
differences in the results between approaches, and when they might be
used interchangeably, is an interesting area for further research.
Regardless of the type of economic model used, or indeed whether there
is effect modification at all, the relevant effectiveness inputs must be
produced appropriately. MAIC and STC cannot produce estimates of
individual event probabilities, and are limited to producing average
event probabilities in the aggregate study population of an indirect
comparison. The population-average marginal treatment effects that these
methods estimate are specific to the distribution of baseline risk (as
well as all covariates) in the aggregate study population, and cannot be
applied to a different baseline risk distribution in a target population
even if the covariates are similar. At present, ML-NMR is the only
population adjustment method that can produce individual or average
event probabilities, and can do so in any target population of
interest.\autocite{Phillippo2020_methods, Phillippo2022_multinomial}

For decisions based purely on effectiveness, when effect modification is
present we propose that decision-makers look at both the
population-average conditional and marginal estimates and their
respective treatment rankings to assess whether these are in conflict
(\cref{sec:recommendations}). If the rankings agree, then there is no
substantial conflict and a single treatment decision may be justified
for the entire population. However, if the rankings are in conflict then
a single treatment decision cannot satisfy both decision questions for
the entire population, and decision-makers may wish to consider
splitting decisions into subgroups. This proposal necessitates using an
analysis method that can produce both conditional and marginal
estimates. For analyses of single trials, this is possible using
standard regression adjustment, followed by the marginalisation approach
of Zhang\autocite{Zhang2008} (for binary outcomes) or Daniel et
al.\autocite{Daniel2020} (for time-to-event outcomes). For
population-adjusted indirect comparisons or evidence syntheses of
multiple studies, current implementations of MAIC and STC cannot produce
estimates of both estimands, and moreover cannot typically produce
estimates for a chosen decision target population. ML-NMR can produce
estimates of both the conditional and marginal estimands in any target
population of interest, as well as the necessary quantities for
cost-effectiveness models such as average event probabilities or
subgroup/individual event probabilities, making this a powerful tool for
analysts and
decision-makers.\autocite{Phillippo2020_methods, Phillippo2022_multinomial}

\section*{Highlights}

\begin{enumerate}
\def\labelenumi{\arabic{enumi}.}
\item
  \textbf{What is already known} \quad When using non-collapsible
  measures of treatment effects, such as odds ratios or hazard ratios,
  marginal and conditional estimands have different interpretations and
  will not generally coincide, even in the absence of effect
  modification. The presence of effect modification means that there may
  not be a single most effective treatment for all individuals or
  subgroups in the population.
\item
  \textbf{What is new} \quad We argue that population-average
  conditional and marginal estimands both quantify average effectiveness
  over a population but correspond to different decision questions,
  either to maximise the average effect for individuals in the
  population, or to minimise (or maximise) average event probabilities
  respectively. When effect modification is present, we show that these
  are no longer aligned and can result in conflicting treatment
  rankings. In such cases, making a single treatment decision can result
  in choosing an inferior treatment for the majority of individuals, or
  one with a worse expected number of events overall.
\item
  \textbf{Potential impact} \quad We provide recommendations for
  decision-making in the presence of effect modification, for decisions
  based both purely on effectiveness and on cost-effectiveness. ML-NMR
  is at present the only population adjustment method that can produce
  the necessary estimates in any target population of interest. Where
  allowable, making decisions by subgroups may result in patients being
  given a more effective treatment for them and result in greater
  cost-effectiveness overall.
\end{enumerate}

\section*{Acknowledgements}

DMP was supported by the UK Medical Research Council, grant numbers
MR/R025223/1 and MR/W016648/1. ARA is employed by Novo Nordisk but
declares no conflicts of interest as this research is purely
methodological.

\appendix

\section{Definitions of estimands in terms of potential
outcomes}\label{definitions-of-estimands-in-terms-of-potential-outcomes}

\label{sec:defs_potential_outcomes}

We can define each of the estimands given in \cref{sec:definitions}
using the potential outcomes framework.\autocite{Rubin1974} Let \(Y(k)\)
denote potential outcomes for individuals receiving treatment \(k\), and
let \(g(\cdot)\) be a suitable link function such as the logit link
function.

The individual-level conditional treatment effects
\eqref{eqn:individual} for an individual with covariates \(\bm{x}\)
receiving treatment \(b\) compared to \(a\) are defined as
\begin{equation}
  \gamma_{ab}(\bm{x}) = g\mleft( \mathbb{E}\mleft(Y(b) \mid \bm{x}\mright) \mright) - g\mleft( \mathbb{E}\mleft(Y(a) \mid \bm{x}\mright) \mright)
\end{equation} where the expectations are over the distribution of
potential outcomes for an individual conditional on the covariates.

The population-average conditional treatment effects
\eqref{eqn:conditional} for treatment \(b\) compared to \(a\) in
population \(P\) are defined as \begin{equation}
  d_{ab(P)} = \mathbb{E}_{(P)}\mleft( g\mleft(\mathbb{E}(Y(b) \mid \bm{x})\mright) \mright) - \mathbb{E}_{(P)}\mleft( g\mleft(\mathbb{E}(Y(a) \mid \bm{x})\mright) \mright)
\end{equation} where the inner expectations are over the distribution of
potential outcomes for an individual conditional on the covariates, and
the outer expectations are over the population \(P\). In other words,
this is the expectation of the individual-level conditional treatment
effects over the population,
\(d_{ab(P)} = \mathbb{E}_{(P)}(\gamma_{ab}(\bm{x}))\).

The population-average marginal treatment effects \eqref{eqn:marginal}
for treatment \(b\) compared to \(a\) in population \(P\) are defined as
\begin{equation}
  \Delta_{ab(P)} = g\mleft( \mathbb{E}_{(P)}(Y(b)) \mright) - g\mleft( \mathbb{E}_{(P)}(Y(a)) \mright)
\end{equation} where the expectations are over the distribution of
potential outcomes in the population \(P\).

\section{Proof that effect modification is necessary for
rank-switching}\label{proof-that-effect-modification-is-necessary-for-rank-switching}

\label{sec:em_proof}

We prove here that rank-switching between the population-average
conditional and marginal effects can only occur in the presence of
effect modification, between treatments that have different interaction
terms (i.e.~no shared effect modifier assumption), and when this causes
treatment ranks to change across individuals/subgroups in the
population. We shall consider the population-average conditional and
marginal estimands \(d_{ab(P)}\) and \(\Delta_{ab(P)}\) for any two
treatments \(a\) and \(b\), and consider without loss of generality that
we have \(d_{ab(P)} > 0\).

Firstly, let us show that when there is no effect modification between
treatments \(a\) and \(b\), there can be no rank switching between the
population-average conditional and marginal estimands for these two
treatments (i.e.~\(d_{ab(P)}\) and \(\Delta_{ab(P)}\) must have the same
sign). Let \(\bm{\beta}_{2,b} - \bm{\beta}_{2,a} = \bm{0}\) so that
there is no effect modification between these two treatments. (Either
there is no effect modification at all so
\(\bm{\beta}_{2,b} = \bm{\beta}_{2,a} = \bm{0}\), or the shared effect
modifier assumption is made for these two treatments so
\(\bm{\beta}_{2,b} = \bm{\beta}_{2,a}\).) In this case, we have that
\begin{equation}
  d_{ab(P)} = \gamma_b - \gamma_a
\end{equation} following the definition of \(d_{ab(P)}\) in equation
\eqref{eqn:conditional}. We have \(d_{ab(P)} > 0\), so therefore
\begin{equation}
  \gamma_b > \gamma_a.
\end{equation} Since \(g(\cdot)\) is monotonically increasing by
definition as a link function, and following the definition of the
individual event probabilities in equation \eqref{eqn:p_individual},
this means that \begin{equation}
  \pi_b(\bm{x}) > \pi_a(\bm{x}) \quad \textrm{ for all } \bm{x},
\end{equation} and therefore \begin{equation}
  \bar{\pi}_b > \bar{\pi}_a
\end{equation} following the definition of the average event
probabilities in equation \eqref{eqn:pbar}. From the definition of
\(\Delta_{ab(P)}\) in \eqref{eqn:marginal}, and again since \(g(\cdot)\)
is monotonically increasing, this means that \begin{equation}
  \Delta_{ab(P)} > 0.
\end{equation} Therefore, if there is no effect modification then
\(d_{ab(P)}\) and \(\Delta_{ab(P)}\) cannot be in conflict; they must
have the same sign and the corresponding treatment ranks must be in
agreement for these two treatments. A similar argument extends this fact
to the case where \(\bm{\beta}_{2,b} - \bm{\beta}_{2,a}\) is non-zero
but \(\bm{x}\) is constant within the population, so that there is no
substantive effect modification within the population.

Secondly, let us show that effect modification is necessary for the two
estimands \(d_{ab(P)}\) and \(\Delta_{ab(P)}\) to be in conflict
(i.e.~to have opposite signs), and that this effect modification must
cause the individual-level treatment effects to change ranks across
individuals/subgroups within the population for the conflict to occur.
Consider still that \(d_{ab(P)} > 0\), but now that
\(\Delta_{ab(P)} < 0\). From the fact that \(g(\cdot)\) is monotonically
increasing, \(\Delta_{ab(P)} < 0\) means that \begin{equation}
  \bar{\pi}_b < \bar{\pi}_a
\end{equation} following definition \eqref{eqn:pbar}. Therefore, there
must be some values of \(\bm{x}\) within the population, say the set
\(\mathfrak{X}^*\), where \begin{equation}
  \pi_b(\bm{x}) < \pi_a(\bm{x}) \quad \textrm{ for } \bm{x} \textrm{ in $\mathfrak{X}^*$},
\end{equation} and therefore where \begin{equation}
  \label{eqn:ind_switch}
  \gamma_{ab}(\bm{x}) = \eta_b(\bm{x}) - \eta_a(\bm{x}) =  g(\pi_b(\bm{x})) - g(\pi_a(\bm{x})) < 0 \quad \textrm{ for } \bm{x} \textrm{ in $\mathfrak{X}^*$},
\end{equation} again using the fact that \(g(\cdot)\) is monotonically
increasing. We also must have that the set \(\mathfrak{X}^*\) does not
constitute the entire population, because if this was the case then
equation \eqref{eqn:ind_switch} implies that
\(\gamma_{ab}(\bar{\bm{x}}) = d_{ab(P)} < 0\) (whereas we have assumed
the contrary). Together we have \(\gamma_{ab}(\bm{x}) < 0\) for
\(\bm{x}\) in \(\mathfrak{X}^*\) and \(\gamma_{ab}(\bm{x}) \ge 0\) for
\(\bm{x}\) not in \(\mathfrak{X}^*\), in other words the individual
treatment ranks change for different individuals/subgroups in the
population, and this necessarily means that \(\gamma_{ab}(\bm{x})\) is
not constant. By definition (equation \eqref{eqn:individual}), this
requires that \(\bm{\beta}_{2,b} - \bm{\beta}_{2,a}\) is non-zero, and
that \(\bm{x}\) is not constant in the population; that is, there is
substantive effect modification between the two treatments in the
population.

We have therefore shown that effect modification is a necessary
condition for rank-switching between the population-average conditional
and marginal effects to occur, and moreover the effect modification must
give rise to different treatment rankings for different
individuals/subgroups in the population if the rank-switching is to
occur.

\printbibliography[heading=bibintoc,title={References}]

\end{document}